\documentclass[sigconf]{acmart}

\usepackage{enumitem}
\usepackage{url}
\usepackage{xcolor}
\usepackage{algorithm}
\usepackage{algorithmic} 
\usepackage{amsmath}
\usepackage{graphicx}
\usepackage{hyperref}
\usepackage{booktabs}
\usepackage{multirow}
\usepackage{makecell}
\usepackage{comment}
\usepackage[utf8]{inputenc}
\usepackage[T1]{fontenc}
\usepackage{soul}
\usepackage{subcaption}

\AtBeginDocument{%
  }

\copyrightyear{2026}
\acmYear{2026}
\setcopyright{cc}
\setcctype{by}
\acmConference[KDD '26]{Proceedings of the 32nd ACM SIGKDD Conference on Knowledge Discovery and Data Mining V.2}{August 09--13, 2026}{Jeju Island, Republic of Korea}
\acmBooktitle{Proceedings of the 32nd ACM SIGKDD Conference on Knowledge Discovery and Data Mining V.2 (KDD '26), August 09--13, 2026, Jeju Island, Republic of Korea}
\acmDOI{10.1145/3770855.3818142}
\acmISBN{979-8-4007-2259-2/2026/08}

\begin{document}

\title{One Sequential Recommendation Model Pretrained from Synthetic Priors Predicts Multiple Datasets}

\author{Woosung Kang}
\email{wskang@kaist.ac.kr}
\orcid{0009-0002-7535-4383}
\affiliation{
  \institution{Korea Advanced Institute of Science \& Technology}
  \city{Daejeon}
  \country{Republic of Korea}
}

\author{Jiwon Jeong}
\email{zzioni@kaist.ac.kr}
\orcid{0009-0005-7535-3104}
\affiliation{
  \institution{Korea Advanced Institute of Science \& Technology}
  \city{Daejeon}
  \country{Republic of Korea}
}

\author{Jonghyeok Shin}
\email{jonghyeok@kaist.ac.kr}
\orcid{0009-0003-7706-8978}
\affiliation{
    \institution{Korea Advanced Institute of Science \& Technology}
    \city{Daejeon}
    \country{Republic of Korea}
}

\author{Jeongwhan Choi}
\email{jeongwhan.choi@kaist.ac.kr}
\orcid{0000-0002-6530-2662}
\affiliation{
  \institution{Korea Advanced Institute of Science \& Technology}
  \city{Daejeon}
  \country{Republic of Korea}
}

\author{Noseong Park}
\authornote{Corresponding author.}
\email{noseong@kaist.ac.kr}
\orcid{0000-0002-1268-840X}
\affiliation{
  \institution{Korea Advanced Institute of Science \& Technology}
  \city{Daejeon}
  \country{Republic of Korea}
}

\renewcommand{\shortauthors}{Woosung Kang, Jiwon Jeong, Jonghyeok Shin, Jeongwhan Choi, and Noseong Park} 

\begin{abstract}
    Existing sequential recommendation models rely on dataset-specific training, where the learned parameters are fitted to the item catalog and the observed interaction distribution of the training data.
    This limits generalization to new domains, typically requiring retraining from scratch.
    In this work, we propose SRPFN, a Prior-data Fitted Network for sequential recommendation --- predicting the next item in a single forward pass without any gradient-based parameter updates in the target domain.
    SRPFN is pretrained offline on 25.6M sequences sampled from a synthetic prior that spans diverse item-to-item transition patterns, learning to produce posterior predictive next-item distributions. 
    At inference time, SRPFN generates recommendations by conditioning on a support set of item-item transition examples from the target domain, adapting to domain-specific patterns without retraining.
    Extensive experiments on five benchmarks across 10 baselines show that SRPFN achieves the best or second-best performance across nearly all metrics and datasets, while being substantially more computationally efficient than trained baselines.
    These results establish that a single model pretrained on synthetic priors can generalize across diverse real-world domains, offering a framework for update-free sequential recommendation.
\end{abstract}



\maketitle
\newcommand\kddavailabilityurl{https://doi.org/10.5281/zenodo.20403804}
\ifdefempty{\kddavailabilityurl}{}{
\begingroup\small\noindent\raggedright\textbf{Resource Availability:}\\
The source code of this paper has been made publicly available at \url{\kddavailabilityurl}.
\endgroup
}

\section{Introduction}

\begin{figure}[!tb]
    \centering
    \begin{subfigure}[t]{0.45\linewidth}
        \centering
        \includegraphics[width=\linewidth]{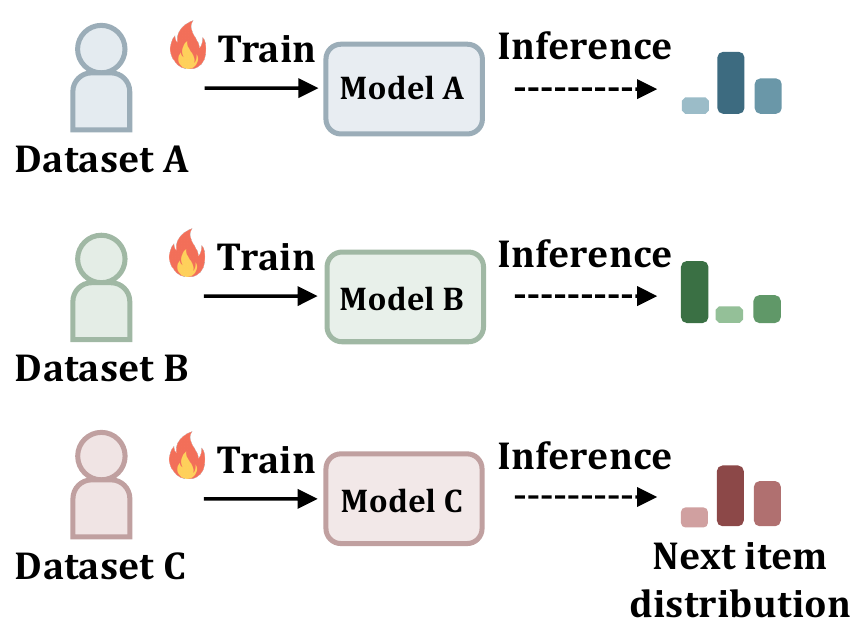}
        \caption{Existing methods}
        \label{fig:teaser:a}
    \end{subfigure}
    \hfill
    \begin{subfigure}[t]{0.54\linewidth}
        \centering
        \includegraphics[width=\linewidth]{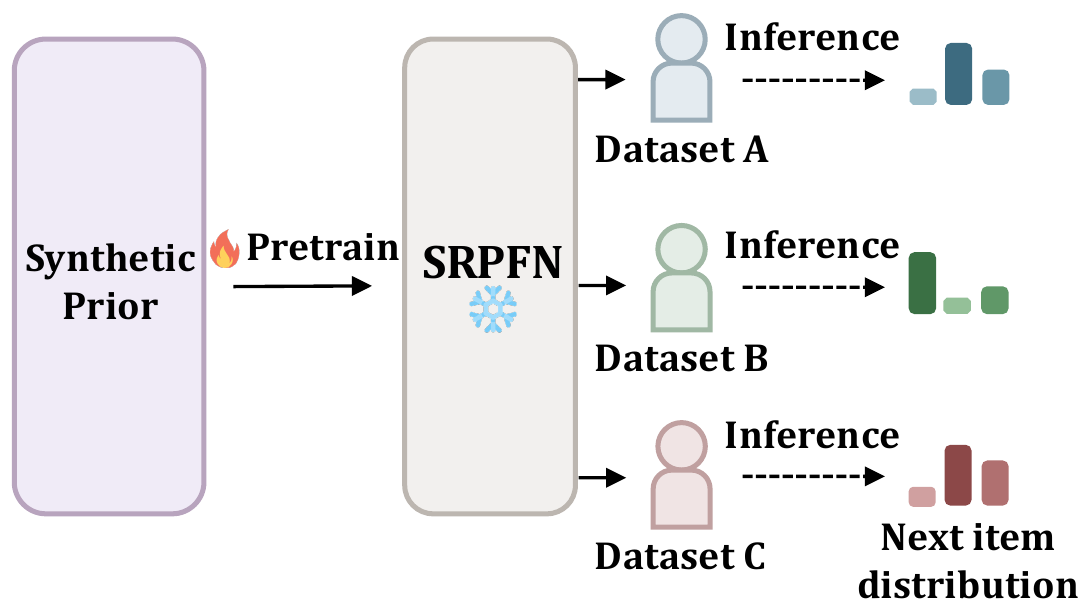}
        \caption{SRPFN}
        \label{fig:teaser:b}
    \end{subfigure}
    \caption{Paradigm comparison. (a) Existing methods train a separate model for each dataset. (b) SRPFN is pretrained once on a synthetic prior and then applied to multiple datasets without retraining --- a setting we refer to as \emph{update-free} inference.}
    \label{fig:teaser}
\end{figure}

Sequential recommendation~\cite{ijcai2019p883,10.1145/3426723} aims to predict the next item that a user will interact with based on their past interaction history.
In practice, item-to-item transition patterns, \emph{i.e.}, the underlying dynamics that govern how users move from one item to the next, vary across datasets~\cite{ren2019repeat, hou2022towards}.
Existing sequential recommendation methods follow a representation learning paradigm~\cite{ijcai2019p883, 10144391}. 
Standard architectures such as GRU4Rec~\cite{Hidasi2015SessionbasedRW} and SASRec~\cite{Kang2018SelfAttentiveSR} rely on fixed-size, ID-based embedding tables defined over a specific item set, with parameters explicitly fitted to the interaction sequences of the training dataset.
However, the transition patterns of a new dataset often differ from those observed during training~\cite{ren2019repeat, hou2022towards}.
Consequently, deploying a trained model on a new dataset typically requires retraining from scratch~\cite{yuan2020peterrec, ijcai2021p639}.
This limitation motivates sequential recommenders that can be applied to new datasets without gradient-based parameter updates --- a setting we refer to as \emph{update-free} inference. 
Fig.~\ref{fig:teaser} contrasts this setting with the conventional dataset-specific training paradigm.

\begin{figure}[t!]
    \centering
    \begin{subfigure}[t]{0.3\columnwidth}
        \centering
        \includegraphics[width=\linewidth]{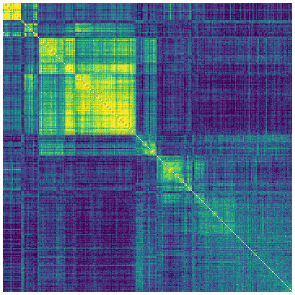}
        \caption{Beauty$^\circ$}
        \label{fig:struct_sim:beauty_sasrec}
    \end{subfigure}
    \hfill
    \begin{subfigure}[t]{0.3\columnwidth}
        \centering
        \includegraphics[width=\linewidth]{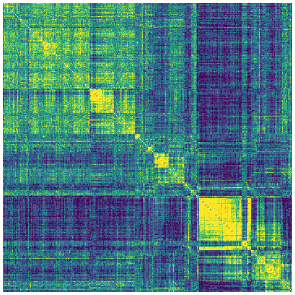}
        \caption{Toys$^\circ$}
        \label{fig:struct_sim:toys_sasrec}
    \end{subfigure}
    \hfill
    \begin{subfigure}[t]{0.3\columnwidth}
        \centering
        \includegraphics[width=\linewidth]{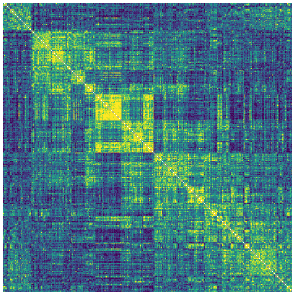}
        \caption{LastFM$^\circ$}
        \label{fig:struct_sim:lastfm_sasrec}
    \end{subfigure}

    \vspace{4pt}

    \begin{subfigure}[t]{0.3\columnwidth}
        \centering
        \includegraphics[width=\linewidth]{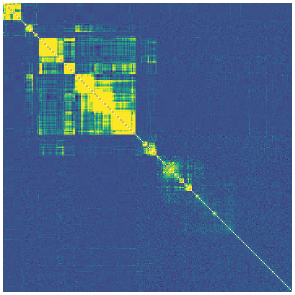}
        \caption{Beauty$^\triangle$}
        \label{fig:struct_sim:beauty_trans}
    \end{subfigure}
    \hfill
    \begin{subfigure}[t]{0.3\columnwidth}
        \centering
        \includegraphics[width=\linewidth]{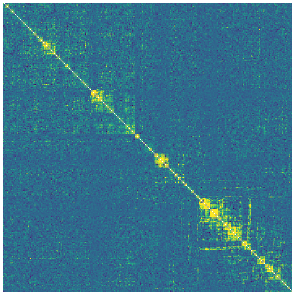}
        \caption{Toys$^\triangle$}
        \label{fig:struct_sim:toys_trans}
    \end{subfigure}
    \hfill
    \begin{subfigure}[t]{0.3\columnwidth}
        \centering
        \includegraphics[width=\linewidth]{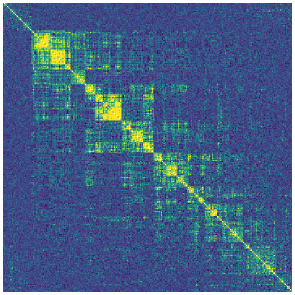}
        \caption{LastFM$^\triangle$}
        \label{fig:struct_sim:lastfm_trans}
    \end{subfigure}

    \caption{
    Structural alignment between SASRec item embeddings ($^\circ$) and low-rank transition statistics ($^\triangle$) across three datasets.
    Each panel shows the item--item cosine similarity matrix within a single embedding space, where brighter colors indicate more similar items.
    }
    \label{fig:struct_sim}
\end{figure}

Previous work has shown that neural collaborative filtering models mainly capture the low-rank structure in interaction data~\cite{steck2019ease, ijcai2024p227}.
We observe a similar phenomenon in the sequential recommendation setting, where the item--item geometry learned by SASRec closely aligns with that induced by a low-rank factorization of transition statistics.
For each dataset, we verify this by comparing $K \times K$ cosine similarity matrices over the top-$K$ most popular items ($K=400$), computed from trained SASRec embeddings and truncated-SVD transition embeddings, as visualized in Fig.~\ref{fig:struct_sim}.

We compare the SASRec and transition-based similarity matrices using centered kernel alignment (CKA)~\cite{kornblith2019cka}.
The observed alignment is high, with CKA values of $0.65$, $0.55$, and $0.62$ on Beauty, Toys, and LastFM, respectively.
These values substantially exceed a conservative null based on randomly initialized embeddings of the same size ($0.23$, $0.39$, and $0.33$), with the gap most pronounced on Beauty and LastFM, and a permutation test further confirms significance ($p \approx 0.005$).
This alignment indicates that simple spectral summaries of transition patterns reflect the structure learned by SASRec embeddings.
Together, these results suggest that such summaries can serve as informative input representations of the underlying low-rank transition structure without requiring gradient-based training on the target domain.

To generalize inference over diverse transition patterns without dataset-specific training, we leverage Prior-data Fitted Networks (PFNs)~\cite{hollmann2023tabpfn, dooley2023forecastpfn, choi2026learning}. By pretraining a Transformer~\cite{vaswani2017attention} on synthetic datasets sampled from a prior, PFNs learn to approximate the posterior predictive distribution in a single forward pass.
Building on this paradigm, we reformulate sequential recommendation as a \emph{conditional inference} problem and propose \textbf{SRPFN}, a Prior-data Fitted Network for sequential recommendation.

Unlike conventional sequential recommenders requiring gradient-based training on each target dataset, SRPFN adapts predictions in an update-free manner by conditioning on a \emph{support set} --- item-to-item transition examples drawn from the target domain's observed interactions.
SRPFN is designed to approximate the posterior predictive distribution (PPD) of the next item, where the user’s history serves as the \emph{query context}, and the support set provides additional information to refine the prediction~\cite{muller2022transformers}.

To enable generalization across diverse transition patterns, SRPFN is pretrained on synthetic datasets sampled from a parametric prior, rather than being fitted to specific real-world datasets.
Concretely, we adopt the hierarchical degree-corrected stochastic block model (hDCSBM)~\cite{PhysRevE.83.016107,PhysRevX.4.011047} to generate item graphs that capture
the hierarchical and categorical structures commonly observed in real-world datasets~\cite{Clauset2008hier,chamberlain2019scalablehyperbolicrecommendersystems}. 
Interaction sequences are then generated through random walks over these graphs, encouraging the model to learn a general-purpose inference mechanism that can adapt to a wide range of transition patterns~\cite{eksombatchai2018pixie, gori2007itemrank}.
At inference time, SRPFN generates recommendations in a single forward pass, without any gradient-based parameter updates on the target domain.

We validate SRPFN through extensive experiments on 
representative sequential recommendation benchmarks.
Despite being pretrained solely on synthetic data, SRPFN achieves strong overall performance, yielding an average improvement of 7.53\% over the second-best method across the evaluated benchmarks and metrics without receiving any gradient updates on the target data. 
Moreover, SRPFN runs inference on a new dataset in approximately one minute, whereas baselines that rely on dataset-specific training require minutes to hours.

Our contributions are summarized as follows:
\begin{itemize}[leftmargin=2em,topsep=0.5em]
    \item We introduce \textbf{SRPFN}, a Prior-data Fitted Network that reformulates sequential recommendation as conditional inference, enabling next-item prediction in a single forward pass without updating parameters for the target domain.
    \item We design a hierarchical stochastic block model-based prior that generates diverse transition patterns, and show that it covers the statistical regimes observed in real-world datasets.
    \item Extensive experiments on five benchmarks demonstrate that SRPFN achieves an average improvement of 7.53\% over the strongest baselines across all metrics, validating update-free inference as a viable paradigm for sequential recommendation.
    \item SRPFN runs inference on a new dataset in approximately one minute, while baselines that require dataset-specific training take about 10 minutes (SASRec) to 15 hours (FEARec), highlighting the efficiency of update-free inference.
\end{itemize}

\section{Related Work}

\subsection{Sequential Recommender Systems}
Sequential recommender systems model user behavior from ordered interaction sequences to predict the next item, typically using ID-based embeddings learned from interaction sequences~\cite{ijcai2019p883,10.1145/3426723}.
Early work is rooted in statistical modeling, including matrix factorization (MF) methods~\cite{hu2008cf, 5197422} that capture collaborative signals from the user--item interaction matrix, but such approaches are limited in modeling short-term intent or fine-grained sequential dependencies.
To address this limitation, Markov chain (MC)--based models explicitly incorporate sequential information by modeling item-to-item dependencies~\cite{10.5555/1795114.1795167, rendle2010fpmc}.

With the advent of deep learning, neural sequential models were proposed to capture more expressive and nonlinear dependency structures in user interaction sequences.
Recurrent neural network-based methods~\cite{Hidasi2015SessionbasedRW, 10.1145/3132847.3132926} model temporal dynamics through recurrent hidden states, while convolutional approaches~\cite{10.1145/3159652.3159656, 10.1145/3289600.3290975} capture local and global sequential patterns.
More recently, Transformer-based architectures have become the dominant paradigm in sequential recommendation.
SASRec~\cite{Kang2018SelfAttentiveSR} introduces causal self-attention to model long-range dependencies, and BERT4Rec~\cite{sun2019bert4rec} adopts bidirectional self-attention with masked item prediction to enhance contextual representation learning.

Building on Transformer backbones, several studies improve robustness and representation quality via contrastive self-supervised objectives and data augmentation strategies~\cite{xie2022CLS4Rec,liu2021coserec,qui2022duorec}.
More recently, a complementary line of work has explored filter-based inductive biases for sequential recommendation, including frequency-domain filtering to mitigate noise or over-smoothing, and convolution-inspired filters that model local and position-dependent temporal patterns~\cite{zhou2022fmlprec, du2023fearec, shin2025tvrec}.
Despite these advances, such methods largely follow a representation learning paradigm, in which model parameters are optimized to dataset-specific transition patterns.

\subsection{Alternatives to Target Domain Training}
The limitations of dataset-specific training have motivated several lines of work aimed at improving generalization in sequential recommendation.
These approaches differ in how they incorporate target domain information and whether they require parameter updates.

A direct alternative is to rely on statistics computed from the target interactions, such as popularity, first-order transitions, item co-occurrence, or nearest-neighbor retrieval~\cite{rendle2010fpmc, 10.1145/963770.963776}.
These methods require no parameter training, but their scoring rules are fixed and dataset-specific.
SRPFN also uses target domain statistics, but only as input evidence to a single pretrained model rather than as fixed scoring rules.

Another line of work retains the standard training paradigm but performs additional optimization at inference time.
Test-Time Training (TTT)~\cite{sun2020ttt, yang2024ttt4rectesttimetrainingapproach}, user-specific optimization frameworks~\cite{xie2025breaking}, and meta-learning approaches~\cite{finn2017maml, lee2019melu} all rely on gradient updates at inference time to adapt the model to a new user or context.
In contrast, SRPFN adapts in a single feed-forward pass conditioned on a support set, without any inference-time optimization.

A separate line of research replaces ID-based embeddings with representations derived from auxiliary modalities.
Frameworks such as ZESRec~\cite{ding2022zeroshot} and UniSRec~\cite{hou2022towards} encode items using pretrained textual or visual encoders, while RecFormer~\cite{li2023text} treats item sequences as text streams.
By mapping items into shared semantic spaces, these approaches enable cross-domain transfer but rely on the availability of side information and may fail to capture domain-specific transition patterns.
SRPFN, by contrast, operates purely on interaction data and conditions predictions on item-to-item transitions observed in the target domain.

Taken together, these alternatives differ in how they use target domain information.
Training-free heuristics use it via fixed scoring rules; inference-time optimization methods use it to update model parameters; and modality-based transfer methods reduce the need for target domain training by relying on auxiliary information.
SRPFN takes a different route, keeping a single set of pretrained parameters fixed while using statistics from the target domain as input evidence for support-set conditioning.

\section{Preliminaries}
\label{sec:prelim}
In this section, we first formulate the standard sequential recommendation task and introduce a posterior predictive learning framework that motivates our synthetic prior-fitting approach.

\subsection{{Problem Formulation}}
\label{sec:prelim:seqrec}
Assume that we have a set of users $\mathcal{U}$ and a set of items $\mathcal{I}$,
where $u \in \mathcal{U}$ denotes a user and $i \in \mathcal{I}$ denotes an item.
The numbers of users and items are $|\mathcal{U}|$ and $|\mathcal{I}|$, respectively.
For each user $u$, we observe a chronologically ordered interaction sequence
$s_u = (i_{u,1}, i_{u,2}, \dots, i_{u,n_u})$, where $n_u = |s_u|$ denotes the sequence length.
Given the prefix $s_{u,1:n_u-1} = (i_{u,1}, \ldots, i_{u,n_u-1})$,
the goal of sequential recommendation is to predict the next item $i_{u,n_u}$.

\subsection{Posterior Predictive Learning with a Synthetic Prior}
\label{sec:prelim:ppd}
We adopt the posterior predictive learning paradigm underlying Prior-data Fitted Networks (PFNs)~\cite{muller2022transformers, hollmann2023tabpfn}. 
We denote the input by $x$ and the corresponding target by $y$, and let $D_{\text{support}}$ denote a support set that provides auxiliary context for predicting $y$ from $x$.
A data-generating mechanism $\phi$ is drawn from a prior distribution $p(\phi)$.
The PPD is then given by
\begin{equation}
\label{eq:ppd}
p(y \mid x, D_{\text{support}})
=
\int p(y \mid x, \phi)\, p(\phi \mid D_{\text{support}})\, d\phi .
\end{equation}

While exact computation of the PPD is generally intractable, PFNs approximate it directly using a neural network $q_{\theta}$, trained offline on synthetic datasets sampled from the prior.
Specifically, a synthetic dataset $D$ is generated by first sampling a mechanism $\phi$ and then drawing samples according to that mechanism.
For each synthetic dataset, the neural predictor $q_{\theta}$ is trained to approximate the PPD by minimizing
\begin{equation}
\label{eq:prior_fit}
\mathcal{L}_{\mathrm{prior}}
=
\mathbb{E}_{D}
\big[
-\log q_{\theta}(y \mid x, D_{\text{support}})
\big].
\end{equation}
Here, $q_{\theta}$ denotes a neural approximation to the PPD.
After pretraining, the model conditions on a support set at inference time and produces predictions in a single forward pass, without any target domain gradient updates.

\section{Methodology}
We now instantiate the posterior predictive framework of Section~\ref{sec:prelim:ppd}.
The data-generating mechanism $\phi$ corresponds to the process of generating item graphs and sampling random walks, as detailed in Section~\ref{sec:methods:syn_data}.
Under this formulation, the input $x$ represents the observed user history prefix $s_{u, 1:n_u-1}$, the target $y$ denotes the next item $i_{u, n_u}$, and the support set $D_{\text{support}}$ comprises transition examples that supply local structural context.
The input representations are derived in Section~\ref{sec:methods:rep}, and the neural approximator $q_\theta$ is implemented by the architecture described in Section~\ref{sec:methods:arch}, which integrates the query history and support set via a cross-attention mechanism.

\subsection{Synthetic Prior}
\label{sec:methods:syn_data}

Motivated by transition patterns commonly observed in real-world sequential data, we design a synthetic prior to capture a representative spectrum of transition dynamics.
Instead of focusing on a single dataset or a fixed transition pattern, we construct a generative process whose induced statistics exhibit diverse behaviors, including varying degrees of popularity skew and transition determinism.
To achieve this, we proceed in two steps: 
\begin{enumerate}[label=\roman*),leftmargin=2em]
    \item constructing a hierarchical item graph to model structural constraints (Section~\ref{sec:methods:syn_data_graph}), and
    \item simulating random walks to generate interaction sequences (Section~\ref{sec:methods:syn_data_rw}).
\end{enumerate} 

\subsubsection{Graph Generation}
\label{sec:methods:syn_data_graph}
To generate synthetic item graphs that reflect the structural properties of a real-world catalog, we adopt a hierarchical degree-corrected stochastic block model (hDCSBM)~\cite{PhysRevE.83.016107, PhysRevX.4.011047}.
We chose this generative model because it naturally captures three key properties observed in real interaction graphs. 
First, it produces hierarchical community structure in which items are organized into macro-communities and nested micro-communities, capturing coarse-to-fine categorical organization (e.g., categories and subcategories).
Second, it induces heterogeneous popularity by drawing item-level in- and out-degrees from a power-law distribution~\cite{Newman_2005}, reproducing the heavy-tailed popularity skew commonly observed in practice~\cite{abdol2017controllingpopbias, Klimashevskaia2024}.
Third, it incorporates core-periphery organization by assigning importance scores to communities, so that high-importance communities exhibit denser internal connectivity and finer micro-granularity~\cite{BORGATTI2000375}.

Under this model, transition probabilities are naturally concentrated within micro- and macro-communities due to the block structure, while cross-macro transitions arise from controlled inter-block connectivity, balancing local coherence with global exploration.
The resulting directed graph is row-normalized to obtain a stochastic transition kernel 
$\mathbf{G} \in \mathbb{R}^{|\mathcal{I}|\times|\mathcal{I}|}$, where $G_{ij}$ denotes the probability of transitioning from item $i$ to item $j$.

\subsubsection{Sequence Generation}
\label{sec:methods:syn_data_rw}
Given the stochastic transition kernel $\mathbf{G}$, we generate a synthetic interaction sequence for each user through random walks following prior work~\cite{eksombatchai2018pixie, gori2007itemrank}.
For user \(u\), the sequence length \(n_u\) is sampled from a power-law distribution, and the initial item \(i_{u,1}\) is drawn proportional to item out-degree~\cite{Newman_2005, yin2012longtail}.
At each step \(t\), we sample a lag \(\ell_t\) from a geometric distribution, clipped to the range \(\{1,\dots,\min(t,L)\}\), to incorporate short-term dependencies.
This procedure retrieves a reference state $\tilde{i}_{u,t} = i_{u,t-\ell_t}$, effectively simulating a user returning to a previous interaction to initiate a new transition, rather than strictly following the immediate predecessor.
The next item is then sampled from the transition probabilities of this reference state:
\begin{align}
i_{u,t+1} \sim \mathrm{Categorical}\!\left(\mathbf{G}_{\tilde{i}_{u,t},:}\right).
\end{align}
By allowing transitions from non-adjacent previous interactions, this mechanism produces sequences that exhibit local coherence when the lag is short and occasional long-range revisits when the lag is large.

\subsubsection{Synthetic Prior Diagnostics}
\begin{figure}[t]
    \centering
    \begin{subfigure}[t]{0.485\columnwidth}
        \centering
        \includegraphics[width=\linewidth]{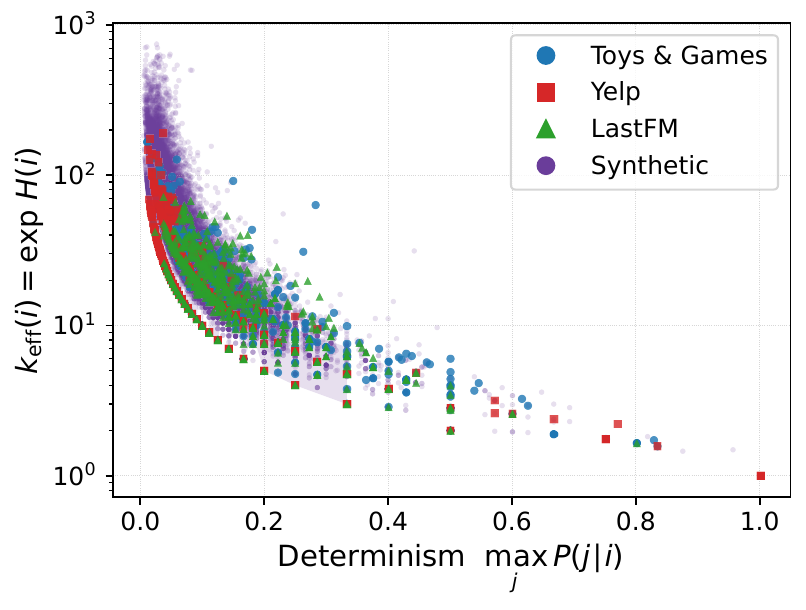}
        \caption{First-order transition pattern.}
        \label{fig:syn_prior_trans}
    \end{subfigure}\hfill
    \begin{subfigure}[t]{0.485\columnwidth}
        \centering
        \includegraphics[width=\linewidth]{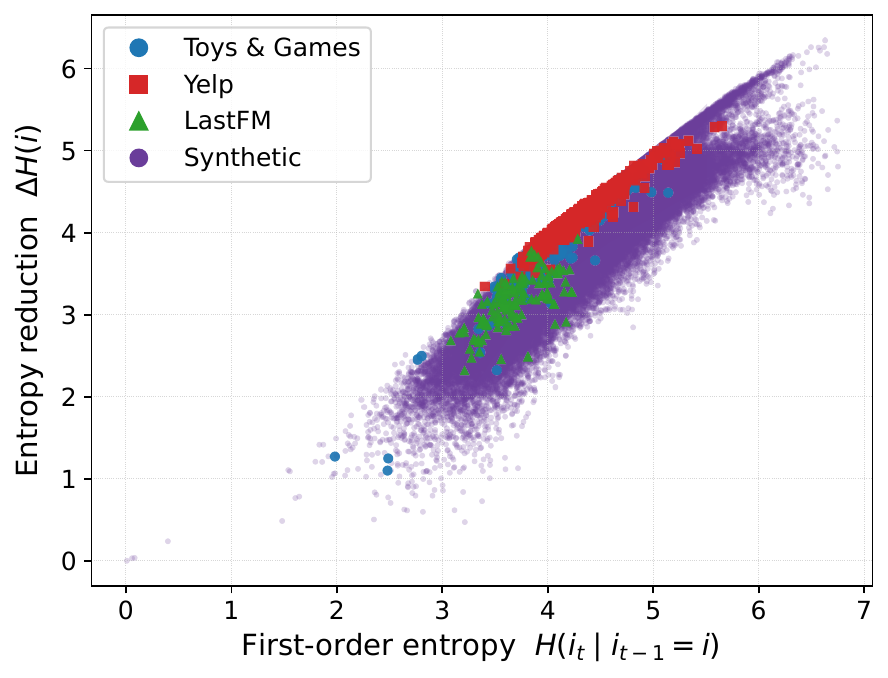}
        \caption{Second-order entropy reduction.}
        \label{fig:syn_prior_ho}
    \end{subfigure}
    \caption{
    Transition diagnostics of the synthetic prior compared with real datasets.
    }
    \label{fig:syn_prior_diagnostics}
\end{figure}

Fig.~\ref{fig:syn_prior_diagnostics} compares the transition statistics induced by the synthetic prior with those observed in real-world datasets.

For the first-order transition structure, we estimate an empirical transition matrix $\hat{\mathbf{P}}$ from consecutive item pairs.
For each source item $i$, we compute the transition determinism
$D(i)=\max_j \hat{P}_{ij}$ and the effective branching factor $k_{\mathrm{eff}}(i)=\exp(H^{(1)}(i))$, where $H^{(1)}(i)$ is the first-order conditional entropy.
As shown in Fig.~\ref{fig:syn_prior_trans}, the synthetic prior spans a broad region in the $(D,k_{\mathrm{eff}})$ plane that overlaps with the regimes occupied by the real datasets.

To assess dependencies beyond first-order transitions, we measure the entropy reduction $\Delta H(i)=H^{(1)}(i)-H^{(2)}(i)$ obtained by conditioning on the previous two items, where $H^{(2)}(i)$ denotes the second-order conditional entropy.
Fig.~\ref{fig:syn_prior_ho} shows that the synthetic prior covers the near-linear relationship between $H^{(1)}(i)$ and $\Delta H(i)$ observed in the real datasets, while also spanning weaker higher-order dependencies.
Overall, these diagnostics suggest that the synthetic prior exposes SRPFN to a diverse range of transition regimes during pretraining.

\subsection{Input Construction}
\label{sec:methods:input}
Given a target domain's observed interaction sequences, we derive low-rank representations that summarize collaborative and transition signals (Section~\ref{sec:methods:rep}), and a support set that provides transition information for each query (Section~\ref{sec:methods:support}).

\subsubsection{Low-rank Representation}
\label{sec:methods:rep}
We construct low-rank representations to summarize collaborative signals and item--item transition patterns~\cite{hu2008cf, rendle2010fpmc}.
To strictly prevent information leakage, we exclude the target item $i_{u,n_u}$ and construct these matrices solely from the observed context $s_{u,1:n_u-1}$.

Formally, we define a user--item interaction matrix \(\mathbf{R} \in \mathbb{R}^{|\mathcal{U}| \times |\mathcal{I}|}\) and an item--item transition matrix \(\mathbf{P} \in \mathbb{R}^{|\mathcal{I}| \times |\mathcal{I}|}\).
The entries of \(\mathbf{R}\) represent the frequency of user interactions:
\begin{equation}
\label{eq:interaction_R}
\mathbf{R}_{ui}
:=
\sum_{t=1}^{n_u-1} \mathbb{I}\!\left[i_{u,t}=i\right].
\end{equation}
For the transition matrix \(\mathbf{P}\), we aggregate multi-hop co-occurrences with inverse-distance weighting to capture sequential dependencies within a local context window:
\begin{equation}
\label{eq:multi_hop_P}
\mathbf{P}_{ij}
:=
\sum_{u \in \mathcal{U}}
\sum_{t=1}^{n_u-1}
\sum_{h=1}^{H}
\frac{1}{h}\,
\mathbb{I}\!\left[
i_{u,t}=i \;\land\; i_{u,t+h}=j \;\land\; t+h < n_u
\right],
\end{equation}
where $H$ is the maximum hop distance and $t \in \{1,\dots,n_u-1\}$ is used to index interactions.

Before factorization, both $\mathbf{R}$ and $\mathbf{P}$ are transformed using Positive Pointwise Mutual Information (PPMI)~\cite{church-hanks-1990-word, Bullinaria2012}, and then factorized through truncated randomized SVD~\cite{halko2010findingstructurerandomnessprobabilistic} to obtain low-rank input representations:
$\mathbf{R}\approx\mathbf{U}^{(R)} \mathbf{\Sigma}^{(R)} (\mathbf{V}^{(R)})^\top$ and $\mathbf{P}\approx\mathbf{U}^{(P)} \mathbf{\Sigma}^{(P)} (\mathbf{V}^{(P)})^\top$.

From these factorizations, we derive four embeddings.
From the interaction factors $\mathbf{E}^{(R)}$, we obtain user--item collaborative embeddings:
$\mathbf{u}_{\mathrm{cf}}(u) := \mathbf{E}^{(R)}_{\text{row}}[u]$ and
$\mathbf{i}_{\mathrm{cf}}(i) := \mathbf{E}^{(R)}_{\text{col}}[i]$, where $\mathbf{E}^{(R)}_{\text{row}} = \mathbf{U}^{(R)} (\mathbf{\Sigma}^{(R)})^{1/2}$ and
$\mathbf{E}^{(R)}_{\text{col}} = \mathbf{V}^{(R)} (\mathbf{\Sigma}^{(R)})^{1/2}$.
From the transition factors $\mathbf{E}^{(P)}$, we obtain source-/destination-side item embeddings:
$\mathbf{i}_{\mathrm{src}}(i) := \mathbf{E}^{(P)}_{\text{row}}[i]$ captures outgoing information, and $\mathbf{i}_{\mathrm{dst}}(i) := \mathbf{E}^{(P)}_{\text{col}}[i]$ captures incoming transitions.

These embeddings serve as the input to both the encoder and the support set construction described below.

\subsubsection{Support-Set Construction}
\label{sec:methods:support}

For each test user~$u$, we construct a support set $D_{\text{support}}$ from the transition matrix~$\mathbf{P}$ (Eq.~\eqref{eq:multi_hop_P}).
Let $i_{u,n_u-1}$ be the last observed item in the prefix $s_{u,1:n_u-1}$.
We sample $|D_{\text{support}}|$ destination items proportional to the transition frequencies of $i_{u,n_u-1}$:
\begin{equation}
\label{eq:support_sampling}
s_m \sim \mathrm{Categorical}\!\left(
\frac{\mathbf{P}_{i_{u,n_u-1},\,\cdot}}
     {\sum_{i' \in \mathcal{I}} \mathbf{P}_{i_{u,n_u-1},\,i'}}
\right),
\quad m = 1,\dots,|D_{\text{support}}|,
\end{equation}
yielding the support set $D_{\text{support}} = \{(i_{u,n_u-1},\, s_m)\}_{m=1}^{|D_{\text{support}}|}$.

\subsection{Model Architecture}
\label{sec:methods:arch}
Unlike conventional sequential recommenders that encode domain statistics into parameters, SRPFN is trained to approximate the PPD in Eq.~\eqref{eq:prior_fit}, where the query history induces a prior representation, the support set provides additional information, and the architecture combines them in a single forward pass.

\subsubsection{Encoder}
\label{sec:methods:encoder}
\begin{figure}[t]
    \centering
    \includegraphics[width=\columnwidth]{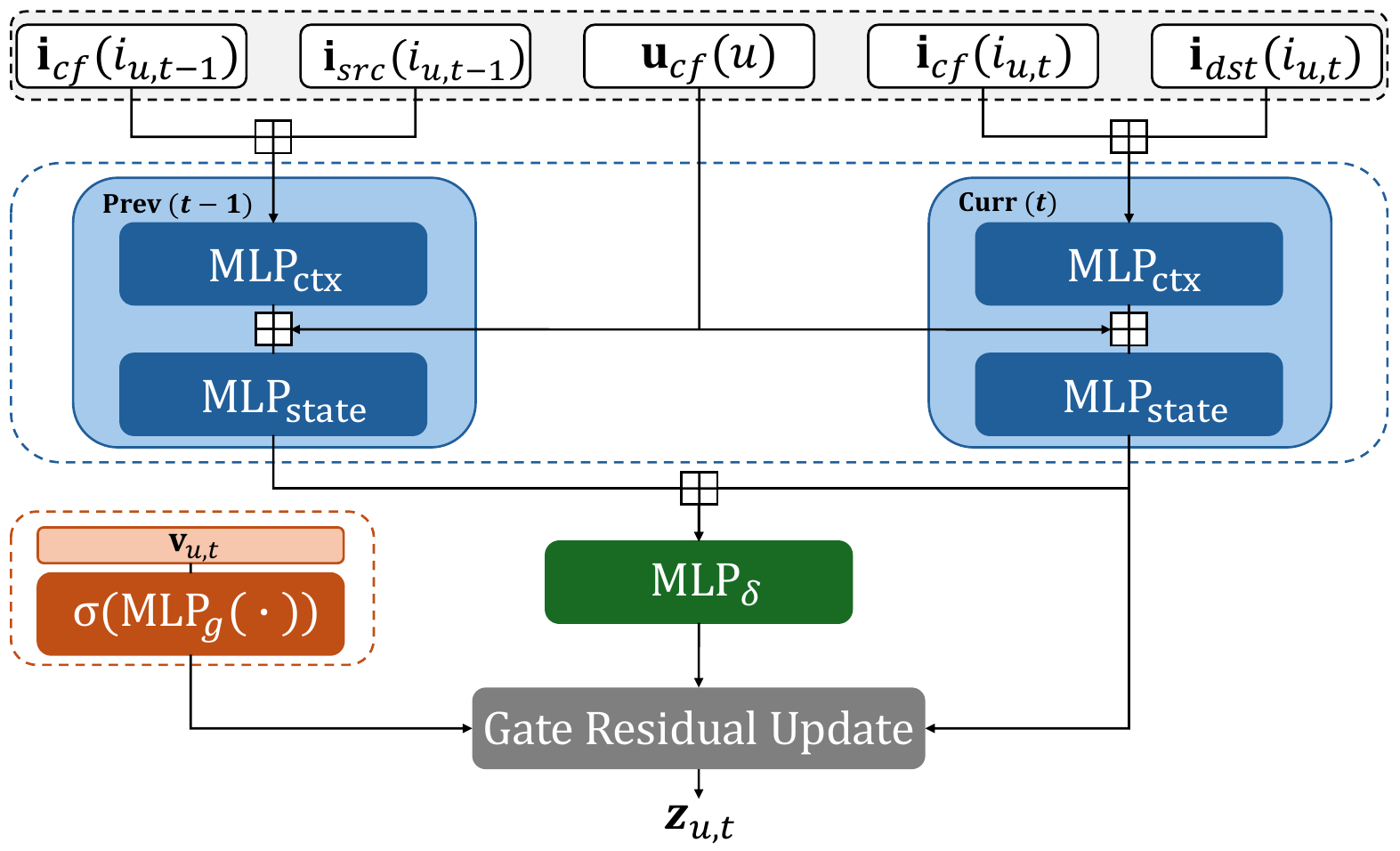}
    \caption{Architecture of the encoder (Section~\ref{sec:methods:encoder}) step $t$. The encoder takes five low-rank embeddings as input: the user collaborative embedding $\mathbf{u}_{\mathrm{cf}}(u)$, the collaborative embeddings of the previous and current items $\mathbf{i}_{\mathrm{cf}}(i_{u,t-1})$ and $\mathbf{i}_{\mathrm{cf}}(i_{u,t})$, and the source- and destination-side transition embeddings $\mathbf{i}_{\mathrm{src}}(i_{u,t-1})$ and $\mathbf{i}_{\mathrm{dst}}(i_{u,t})$.
    $\boxplus$ denotes feature concatenation.}
    \label{fig:encoder}
\end{figure}
We utilize the encoder architecture illustrated in Fig.~\ref{fig:encoder} to produce Transformer input representations.
Rather than relying on absolute coordinates, the encoder emphasizes pairwise signals across consecutive interactions.
For each user $u$, we construct encoder outputs for each step $t \in \{1, \ldots, n_u-1\}$ within the prefix.
For $t \ge 2$, the encoder processes the consecutive pair $(i_{u,t-1}, i_{u,t})$.
For $t = 1$, the preceding interaction is not available, so we zero-pad the source-side context and compute $\mathbf{m}_{u,1}$ accordingly.
We first construct context vectors for the two interactions through a shared projection layer:
\begin{align}    
\begin{split}    
\mathbf{m}_{u,t-1}
&= \mathrm{MLP}_{\mathrm{ctx}}\!\left([\mathbf{i}_{\mathrm{cf}}(i_{u,t-1});\,\mathbf{i}_{\mathrm{src}}(i_{u,t-1})]\right),\\
\mathbf{m}_{u,t}
&= \mathrm{MLP}_{\mathrm{ctx}}\!\left([\mathbf{i}_{\mathrm{cf}}(i_{u,t});\,\mathbf{i}_{\mathrm{dst}}(i_{u,t})]\right).
\end{split}
\end{align}
Each context vector is then combined with the user embedding to form an interaction state:
\begin{align}    
\begin{split} 
\mathbf{e}_{u,t-1}
&= \mathrm{MLP}_{\mathrm{state}}\!\left([\mathbf{u}_{\mathrm{cf}}(u);\,\mathbf{m}_{u,t-1}]\right),
\\
\mathbf{e}_{u,t}
&= \mathrm{MLP}_{\mathrm{state}}\!\left([\mathbf{u}_{\mathrm{cf}}(u);\,\mathbf{m}_{u,t}]\right).
\end{split}
\end{align}
Here, $\mathbf{i}_{\mathrm{src}}(i)$ captures the outgoing information of item $i$ when it appears as a preceding item in a sequence, whereas $\mathbf{i}_{\mathrm{dst}}(i)$ captures the incoming information when it appears as a succeeding item. This formulation allows the encoder to relate the consecutive pair $(i_{u,t-1}, i_{u,t})$ while reflecting the underlying transition patterns~\cite{he2017transrec}.
To model the change between the two interactions, we compute an update vector
\begin{align}    
\boldsymbol{\delta}_{u,t}
= \mathrm{MLP}_{\delta}\!\left([\mathbf{e}_{u,t-1};\mathbf{e}_{u,t}]\right),
\end{align}
and obtain the encoder output via a gated residual update~\cite{srivastava2015highwaynetworks}
\begin{align}    
\mathbf{z}_{u,t}
= \mathrm{LN}\!\left(\mathbf{e}_{u,t}
+ g^{\mathrm{enc}}_{u,t}\cdot \boldsymbol{\delta}_{u,t}\right).
\end{align}
Here, the scalar gate is computed as
\begin{align}    
g^{\mathrm{enc}}_{u,t}=\sigma\!\left(\mathrm{MLP}_g(\mathbf{v}_{u,t})\right),
\end{align}
where $\mathbf{v}_{u,t}$ contains cosine-similarity features among the collaborative and transition embeddings of the consecutive pair $(i_{u,t-1}, i_{u,t})$, including collaborative similarity, source--destination transition similarity, and within-item collaborative--transition consistency.
$\sigma(\cdot)$ denotes the sigmoid function, which constrains the gate value to $(0,1)$.
These features provide a compact consistency signal to modulate the local update.
The standard Layer Normalization $\mathrm{LN}(\cdot)$ is applied for numerical stability before feeding the representations to the Transformer.

\subsubsection{Transformer with Cross-Attention}
\label{sec:methods:transformer}
Given the user's observed prefix $x = s_{u,1:n_u-1}$, we first obtain a sequence of interaction states $\mathbf{Z}_u = [\mathbf{z}_{u,1},\ldots,\mathbf{z}_{u,n_u-1}]$ using the encoder introduced in Section~\ref{sec:methods:encoder}. 
We process this sequence with a causal Transformer backbone to capture temporal dependencies~\cite{vaswani2017attention}.
Let $\mathbf{h}_{u,n_u-1}$ denote the contextualized hidden state at the last position of the prefix.
From this state, we derive a sequential representation $\mathbf{q}_{u,n_u-1}$ via a projection head:
\begin{equation}
\mathbf{q}_{u,n_u-1} = \mathrm{Norm}\!\left(\mathbf{W}_{\mathrm{head}}\,\mathbf{h}_{u,n_u-1}\right),
\end{equation}
where $\mathbf{W}_{\mathrm{head}} \in \mathbb{R}^{d \times d}$ is a learnable projection matrix and
$\mathrm{Norm}(\cdot)$ denotes $\ell_2$-normalization.

To integrate transition information, we encode the support set $D_{\text{support}}$ (Section~\ref{sec:methods:support}) into a memory bank $\mathbf{M}_u \in \mathbb{R}^{|D_{\text{support}}| \times d}$ and apply Multi-Head Cross-Attention (MHCA)~\cite{vaswani2017attention}.
The hidden state $\mathbf{h}_{u,n_u-1}$ serves as the query, while $\mathbf{M}_u$ acts as both keys and values:
\begin{equation}
\mathbf{c}_{u,n_u-1}
=
\mathrm{MHCA}\!\left(
Q = \mathrm{LN}(\mathbf{h}_{u,n_u-1}),\;
K = \mathbf{M}_u,\;
V = \mathbf{M}_u
\right).
\end{equation}
The aggregated context $\mathbf{c}_{u,n_u-1}$ is then mapped to a residual update vector:
\begin{equation}
\boldsymbol{\psi}_{u,n_u-1} = \mathbf{W}_{\mathrm{post}}\,\mathrm{LN}(\mathbf{c}_{u,n_u-1}).
\end{equation}

Finally, we dynamically modulate the influence of the retrieved support information with a learnable scalar gate
$g^{\mathrm{trans}}_{u,n_u-1} \in (0,1)$, computed from the sequential representation:
\begin{equation}
g^{\mathrm{trans}}_{u,n_u-1} = \sigma\!\left(f_{\mathrm{gate}}(\mathbf{q}_{u,n_u-1})\right),
\end{equation}
where $f_{\mathrm{gate}}$ is a lightweight MLP.
Hence, the final representation is obtained by a gated residual update followed by $\ell_2$ normalization:
\begin{equation}
\hat{\mathbf{y}}_{u,n_u-1}
=
\mathrm{Norm}\!\left(
\mathbf{q}_{u,n_u-1}
+
g^{\mathrm{trans}}_{u,n_u-1} \cdot \boldsymbol{\psi}_{u,n_u-1}
\right).
\end{equation}
This mechanism allows the model to incorporate the support information adaptively and can be viewed as an amortized posterior update~\cite{muller2022transformers} in which $D_{\text{support}}$ refines the sequential representation in a single forward pass.

\subsection{Training and Inference}
\label{sec:method:train_infer}
\subsubsection{Training Objective}
During pretraining, the model is trained on synthetic datasets generated from the prior over transition structures, following the posterior predictive formulation in Section~\ref{sec:prelim:ppd}.
For each instance $(x,y,D_{\text{support}})$, we construct a candidate set $\mathcal{C}(x)$ that contains the ground-truth next item $y$ and sampled negatives~\cite{Kang2018SelfAttentiveSR}, and optimize a softmax cross-entropy loss:
\begin{equation}
\label{eq:ce_obj}
\mathcal{L}
=
-\mathbb{E}_{(x,y,D_{\text{support}})}
\left[
\log
\frac{\exp\big(s_{\theta}(y \mid x, D_{\text{support}})\big)}
{\sum\limits_{j \in \mathcal{C}(x)} \exp\big(s_{\theta}(j \mid x, D_{\text{support}})\big)}
\right].
\end{equation}
The matching score is defined as an inner product between the posterior representation $\hat{\mathbf{y}}_{u,n_u-1}$ and the encoded candidate representation.
We interpret the encoding of a candidate item as the encoder output that would be obtained if the item were observed at the next position $n_u$.
Specifically, each candidate item $j \in \mathcal{C}(x)$ is encoded using the same interaction encoder, yielding a next-step representation
$\mathbf{z}_{u,j} \in \mathbb{R}^{d}$.
We then define
\begin{equation}
\label{eq:score_innerprod}
s_{\theta}(j \mid x, D_{\mathrm{support}})
=
\hat{\mathbf{y}}_{u,n_u-1}^{\top}\mathbf{z}_{u,j}.
\end{equation}
This objective corresponds to minimizing the negative log-likelihood in Eq.~\eqref{eq:prior_fit} under a candidate-set restriction.
Under uniform negative sampling, the global minimizer of Eq.~\eqref{eq:ce_obj} recovers the Bayesian PPD of Eq.~\eqref{eq:ppd} renormalized over $\mathcal{C}(x)$, so the ranking induced by the optimal model is consistent with the Bayes-optimal prediction within the candidate set.

\subsubsection{Inference Procedure}
\label{sec:method:train_infer:infer}
At inference time, SRPFN is applied to a target dataset without any gradient updates.
Given the target domain's training interactions, we first construct the low-rank embeddings (Section~\ref{sec:methods:rep}) and the support set for each test user (Section~\ref{sec:methods:support}).
For each test user~$u$, the observed prefix $s_{u,1:n_u-1}$ is then encoded through the Transformer backbone, and the cross-attention mechanism integrates these signals to produce the posterior representation~$\hat{\mathbf{y}}_{u,n_u-1}$.
Candidates are ranked by the inner-product score in Eq.~\eqref{eq:score_innerprod}.

\section{Experiments}
\label{sec:exp}
\subsection{Experimental Setup}
\label{sec:exp:setup}
\subsubsection{Datasets}
We evaluate our model on five public sequential recommendation datasets spanning different application domains.

\begin{itemize}[leftmargin=*, topsep=2pt, itemsep=1.5pt]
  \item \textbf{Amazon Beauty, Sports, and Toys:} obtained from the Amazon review corpus~\cite{10.1145/2766462.2767755}; we select three product categories (Beauty, Sports and Outdoors, and Toys and Games).
  \item \textbf{Yelp}\footnote{\url{https://business.yelp.com/data/resources/open-dataset/}}: a business recommendation dataset; we keep only interactions after \textit{January 1st, 2019}.
  \item \textbf{LastFM}\footnote{\url{https://grouplens.org/datasets/hetrec-2011/}}: a music recommendation dataset containing user--artist interactions, which has relatively long interaction sequences and a small item catalog.
\end{itemize}
We follow the preprocessing protocols in prior work~\cite{10.1145/3340531.3411954, zhou2022fmlprec}.
For all datasets, we group interactions by user, sort them by timestamp in ascending order, and apply 5-core filtering, retaining users and items with at least 5 interactions. Dataset statistics after preprocessing are summarized in Appendix~\ref{app:dataset_stats}.

\subsubsection{Baselines}
We compare SRPFN against two groups of baselines: \emph{trained-on-target} deep sequential recommenders and \emph{training-free} heuristic methods.
For clarity, we note that SRPFN belongs to a third category: \emph{update-free} methods, which may involve offline pretraining but perform inference without any gradient updates on the target domain.

\paragraph{Trained-on-target baselines.}
These models adapt to each target dataset via supervised optimization on the target training split.
We include six representative ID-based sequential recommenders:
GRU4Rec~\cite{Hidasi2015SessionbasedRW},
Caser~\cite{10.1145/3159652.3159656},
SASRec~\cite{Kang2018SelfAttentiveSR},
FMLPRec~\cite{zhou2022fmlprec},
CoSeRec~\cite{liu2021coserec},
and FEARec~\cite{du2023fearec}.

\paragraph{Training-free baselines.}
These baselines rely only on empirical statistics or nearest-neighbor retrieval computed from the target training split, without any parameter updates:
PopRec, Markov-1, Co-Occur, and ItemKNN~\cite{10.1145/963770.963776}.

\subsubsection{Evaluation Settings}
In line with prior work \cite{sun2019bert4rec, 10.1145/3397271.3401111}, we adopt the standard \emph{leave-one-out} evaluation protocol. 
We truncate each user sequence to the 50 most recent interactions.
For each user interaction sequence, the final interaction is reserved for testing, the second-to-last one is used for validation, and all preceding interactions are utilized for training. 
Model performance is measured using top-$k$ Hit Ratio (HR@$k$) and Normalized Discounted Cumulative Gain (NDCG@$k$), and we report HR@$\{1,5,10\}$ and NDCG@$\{5,10\}$.
Following established practices \cite{10.1145/3209978.3210017, 10.1145/3340531.3411954}, we construct the candidate set by pairing the ground-truth item with 99 randomly sampled negative items. 
All metrics are computed based on the rank of the ground-truth item within the candidate set and averaged across all test users. For SRPFN, $|D_{\mathrm{support}}|=16$ was used.

\begin{table*}[t]
\centering
\caption{Main results on five benchmarks.
We report top-$k$ ranking metrics for standard trained-on-target sequential recommenders and training-free baselines, along with SRPFN (evaluated with $|D_{\mathrm{support}}|=16$).
The best and second-best results in each row are denoted in bold and underlined, respectively.
\textit{$\dagger$}: baselines trained on the target domain training split.
\textit{$\ddagger$}: training-free baselines computed from the target domain training split without gradient updates.}
\label{tab:main}
\setlength{\tabcolsep}{4.0pt}
\small
\begin{tabular}{ll cccccc cccc c}
\toprule
\multirow{2}{*}{Dataset} &
\multirow{2}{*}{Metric} &
\multicolumn{6}{c}{Trained-on-target $\dagger$} &
\multicolumn{4}{c}{Training-free $\ddagger$} &
\multicolumn{1}{c}{Update-free} \\
\cmidrule(lr){3-8}\cmidrule(lr){9-12}\cmidrule(lr){13-13}
& &
GRU4Rec & Caser & SASRec & FMLPRec & CoSeRec & FEARec &
PopRec & Markov-1 & Co-Occur & ItemKNN &
SRPFN \\
\midrule

Beauty & HR@1   & 0.1292 & 0.1337 & 0.1870 & 0.1937 & 0.1995 & \underline{0.2140} & 0.0525 & 0.0987 & 0.1330 & 0.1234 & \textbf{0.2364} \\
               & HR@5    & 0.3182 & 0.3032 & 0.3741 & 0.3724 & 0.3821 & \underline{0.3969} & 0.1742 & 0.1704 & 0.2506 & 0.2446 & \textbf{0.4418} \\
               & NDCG@5  & 0.2272 & 0.2219 & 0.2848 & 0.2876 & 0.2951 & \underline{0.3100} & 0.1127 & 0.1349 & 0.1942 & 0.1860 & \textbf{0.3455} \\
               & HR@10   & 0.4207 & 0.3942 & 0.4696 & 0.4609 & 0.4799 & \underline{0.4879} & 0.2899 & 0.2476 & 0.3230 & 0.3213 & \textbf{0.5394} \\
               & NDCG@10 & 0.2601 & 0.2512 & 0.3156 & 0.3162 & 0.3267 & \underline{0.3394} & 0.1498 & 0.1596 & 0.2174 & 0.2106 & \textbf{0.3770} \\
\cmidrule(lr){1-13}

Toys   & HR@1   & 0.1151 & 0.1114 & 0.1878 & 0.1975 & 0.1944 & \underline{0.2185} & 0.0479 & 0.0932 & 0.1275 & 0.1219 & \textbf{0.2371} \\
               & HR@5    & 0.3132 & 0.2614 & 0.3682 & 0.3677 & 0.3633 & \underline{0.3962} & 0.1532 & 0.1626 & 0.2275 & 0.2220 & \textbf{0.4305} \\
               & NDCG@5  & 0.2167 & 0.1885 & 0.2820 & 0.2865 & 0.2821 & \underline{0.3117} & 0.1011 & 0.1275 & 0.1794 & 0.1733 & \textbf{0.3389} \\
               & HR@10   & 0.4256 & 0.3540 & 0.4663 & 0.4581 & 0.4659 & \underline{0.4882} & 0.2403 & 0.2412 & 0.2992 & 0.2980 & \textbf{0.5282} \\
               & NDCG@10 & 0.2530 & 0.2183 & 0.3136 & 0.3156 & 0.3153 & \underline{0.3413} & 0.1291 & 0.1526 & 0.2024 & 0.1977 & \textbf{0.3704} \\
\cmidrule(lr){1-13}

Sports & HR@1   & 0.1127 & 0.1135 & 0.1455 & 0.1501 & 0.1634 & \underline{0.1764} & 0.0658 & 0.0690 & 0.1072 & 0.0874 & \textbf{0.2036} \\
               & HR@5    & 0.3069 & 0.2866 & 0.3466 & 0.3435 & 0.3608 & \underline{0.3743} & 0.2016 & 0.1528 & 0.2127 & 0.2055 & \textbf{0.4297} \\
               & NDCG@5  & 0.2122 & 0.2020 & 0.2497 & 0.2500 & 0.2655 & \underline{0.2788} & 0.1342 & 0.1107 & 0.1615 & 0.1477 & \textbf{0.3206} \\
               & HR@10   & 0.4259 & 0.4014 & 0.4622 & 0.4583 & 0.4725 & \underline{0.4905} & 0.3036 & 0.2381 & 0.2905 & 0.2885 & \textbf{0.5512} \\
               & NDCG@10 & 0.2506 & 0.2390 & 0.2869 & 0.2870 & 0.3015 & \underline{0.3163} & 0.1670 & 0.1380 & 0.1864 & 0.1744 & \textbf{0.3598} \\
\cmidrule(lr){1-13}

Yelp   & HR@1   & 0.1810 & 0.2188 & 0.2375 & 0.2545 & \underline{0.2788} & 0.2576 & 0.0765 & 0.0535 & 0.1433 & 0.1309 & \textbf{0.2891} \\
               & HR@5    & 0.5243 & 0.5111 & 0.5745 & 0.5756 & \underline{0.6052} & 0.5899 & 0.2280 & 0.1312 & 0.2775 & 0.2874 & \textbf{0.6354} \\
               & NDCG@5  & 0.3569 & 0.3696 & 0.4113 & 0.4210 & \underline{0.4486} & 0.4296 & 0.1536 & 0.0916 & 0.2145 & 0.2127 & \textbf{0.4685} \\
               & HR@10   & 0.6990 & 0.6661 & 0.7265 & 0.7241 & \underline{0.7599} & 0.7517 & 0.3405 & 0.2163 & 0.3467 & 0.3657 & \textbf{0.7924} \\
               & NDCG@10 & 0.4137 & 0.4198 & 0.4474 & 0.4692 & \underline{0.4988} & 0.4821 & 0.1898 & 0.1188 & 0.2367 & 0.2380 & \textbf{0.5195} \\
\cmidrule(lr){1-13}

LastFM & HR@1   & 0.0853 & 0.0899 & \underline{0.1211} & 0.0890 & 0.0881 & 0.1064 & 0.0688 & 0.0550 & 0.1138 & 0.1110 & \textbf{0.1257} \\
               & HR@5    & 0.2514 & 0.2982 & \textbf{0.3385} & 0.2642 & 0.2431 & 0.3018 & 0.1881 & 0.1358 & 0.2862 & 0.2706 & \underline{0.3330} \\
               & NDCG@5  & 0.1694 & 0.1960 & \textbf{0.2330} & 0.1770 & 0.1658 & 0.2059 & 0.1289 & 0.0966 & 0.2014 & 0.1914 & \underline{0.2297} \\
               & HR@10   & 0.3972 & 0.4431 & \underline{0.4706} & 0.3697 & 0.3596 & 0.4330 & 0.2963 & 0.2229 & 0.3927 & 0.3798 & \textbf{0.4734} \\
               & NDCG@10 & 0.2163 & 0.2428 & \textbf{0.2775} & 0.2110 & 0.2032 & 0.2483 & 0.1634 & 0.1240 & 0.2359 & 0.2269 & \underline{0.2750} \\
\bottomrule
\end{tabular}
\end{table*}

\subsubsection{Implementation Details}
All experiments are conducted on a single \texttt{NVIDIA RTX A6000} GPU.
For trained-on-target baselines, we use the official implementations released by the original authors and follow their recommended settings.
We set the embedding dimension to 256 for SRPFN and all baselines to ensure consistent model capacity.
All training-free baselines are computed directly from the target training split without any parameter learning.
Additional implementation details are provided in Appendix~\ref{app:baselines}.

Unlike the trained-on-target baselines, SRPFN operates in an update-free manner on the target domain. It is exclusively pretrained on synthetic datasets generated from our prior.

\subsection{Main Results}

Table~\ref{tab:main} reports the main next-item ranking results on five benchmarks, where we compare SRPFN against two groups of baselines: sequential recommenders trained directly on data from the target domain~($\dagger$), and training-free baselines that operate on target domain statistics without gradient-based optimization~($\ddagger$).
This table assesses whether a single model pretrained solely on synthetic data can achieve competitive next-item ranking performance without target domain gradient updates.

Across all datasets and metrics, SRPFN consistently outperforms all training-free baselines by a substantial margin, demonstrating that inference-time conditioning on a support set provides significantly richer signals than fixed statistics such as popularity or first-order transitions alone.
Notably, SRPFN achieves strong overall performance relative to models trained directly on the target domain, yielding an average improvement of 7.53\% over the second-best method across the evaluated benchmarks and metrics, despite not receiving any gradient updates on the target data.

SRPFN ranks first on Beauty, Sports, Toys, and Yelp across all reported metrics.
On LastFM, SRPFN achieves the highest HR@1 and HR@10 while remaining competitive on other metrics, trailing SASRec by small margins on HR@5, NDCG@5, and NDCG@10.
We note that LastFM is characterized by long interaction sequences (average length 48.2) and a relatively compact item catalog ($|\mathcal{I}|{=}3{,}646$), a regime in which dataset-specific training can be particularly effective. 
One possible factor is that applying the PPMI transform to $\mathbf{R}$ and $\mathbf{P}$ emphasizes more balanced co-occurrence patterns, which may make fine-grained sequential nuances less prominent than in models trained directly on this dataset.

The consistent gap between SRPFN and Markov-1 highlights the importance of contextualizing transition information.
Both methods utilize transition statistics from the target domain as their primary signal; however, Markov-1 simply ranks candidates by the maximum first-order transition probability without contextualizing or aggregating multiple transition signals.
In contrast, SRPFN's cross-attention mechanism integrates multiple transition examples from the support set in a context-dependent manner, enabling it to capture higher-order and compositional patterns that first-order counts alone cannot represent.

Overall, these results support that a single model pretrained on synthetic transition patterns can adapt to diverse real-world domains at inference time, achieving performance that typically requires dataset-specific optimization.

\subsection{Generalization to Unseen Categories}
\label{sec:exp:update_free}

Table~\ref{tab:update_free} reports an update-free evaluation on four unseen Amazon categories: Industrial \& Scientific (Ind.\ \& Sci.), Musical Instruments (Instr.), Arts, Crafts \& Sewing (Arts), and Office Products (Office).
All datasets are derived from the Amazon Review Corpus~\cite{10.1145/2766462.2767755}.
All methods are evaluated under an update-free protocol, where no target domain gradient updates or fine-tuning are allowed at test time.
To avoid confounding effects from pretraining overlap, the target categories are selected to be disjoint from those used to pretrain UniSRec and RecFormer.
We follow the evaluation setting described in Section~\ref{sec:exp:setup} and report HR@1 as the primary metric.

\begin{table}[t]
\centering
\caption{Update-free evaluation on unseen Amazon categories (HR@1). The best and second-best methods are denoted in bold and underlined fonts, respectively.}
\label{tab:update_free}
\small
\begin{tabular}{lcccc}
\toprule
Method & Ind.\ \& Sci. & Instr. & Arts & Office \\
\midrule
PopRec   & 0.0986 & \underline{0.1404} & 0.1023 & 0.1605 \\
Markov-1 & 0.0917 & 0.1068 & 0.1329 & 0.1357 \\
Co-Occur & 0.0947 & 0.1264 & 0.1417 & 0.1433 \\
ItemKNN  & \underline{0.1087} & 0.1225 & \underline{0.1722} & \underline{0.1658} \\
\cmidrule(lr){1-5}
RecFormer & 0.0450 & 0.0417 & 0.0688 & 0.0546 \\
UniSRec   & 0.0989 & 0.0871 & 0.1141 & 0.1091 \\
\cmidrule(lr){1-5}
\textbf{SRPFN} & \textbf{0.1995} & \textbf{0.2500} & \textbf{0.3038} & \textbf{0.2654} \\
\bottomrule
\end{tabular}
\end{table}

Under this constraint, standard target-trained sequential recommenders are not applicable, as their scoring depends on item representations learned from a fixed target domain vocabulary.
We compare only approaches that remain well-defined without target domain training:
\begin{itemize}[leftmargin=*, topsep=2pt, itemsep=1.5pt]
  \item \textbf{Statistics-based:} PopRec, Markov-1, Co-Occur, and ItemKNN, computed directly from the target training logs.
  \item \textbf{Pretrained:} UniSRec~\cite{hou2022towards} and RecFormer~\cite{li2023text}, applied using released checkpoints without any target domain adaptation.
\end{itemize}

The results in Table~\ref{tab:update_free} show that simple statistics-based methods remain competitive across categories.
In particular, PopRec achieves non-trivial HR@1 by exploiting marginal item frequency alone, and Markov-1 further benefits from lightweight first-order transition patterns estimated from the target logs.
In contrast, pretrained sequential models often underperform these simple baselines, indicating that their pretrained text-based representations do not capture the target domain's transition patterns without fine-tuning.
Unlike these models, SRPFN does not rely on fixed pretrained item representations; instead, it derives its input from the target domain's own interaction statistics (Section~\ref{sec:methods:rep}).
SRPFN consistently achieves the highest HR@1 across all four categories.
By conditioning predictions on a support set, SRPFN adapts its ranking behavior to target domain characteristics at inference time without requiring parameter updates.
These results show that SRPFN generalizes to unseen categories by combining synthetic-prior pretraining with target domain evidence provided as input.

\subsection{Popularity Bias Analysis}

\begin{table}[t]
\small
\centering
\caption{Results of the popularity bias analysis. Test items are
stratified into Head, Mid, and Tail (20/60/20) by training-set
interaction frequency, and we report group-wise HR@5. The best
performance in each row is in bold.}
\label{tab:pop_bias}
\begin{tabular}{llccc}
\toprule
Dataset & Group & SASRec & FEARec & SRPFN \\
\midrule
\multirow{3}{*}{Beauty}
 & Head & 0.4913 & 0.6312 & \textbf{0.6624} \\
 & Mid  & 0.2232 & 0.2293 & \textbf{0.2799} \\
 & Tail & 0.1198 & 0.1260 & \textbf{0.1959} \\
\cmidrule(lr){1-5}
\multirow{3}{*}{Sports}
 & Head & 0.4819 & 0.6111 & \textbf{0.6726} \\
 & Mid  & 0.1869 & 0.2232 & \textbf{0.2567} \\
 & Tail & 0.0743 & 0.0785 & \textbf{0.1425} \\
\cmidrule(lr){1-5}
\multirow{3}{*}{Yelp}
 & Head & 0.6373 & 0.7871 & \textbf{0.8080} \\
 & Mid  & 0.3757 & 0.4561 & \textbf{0.5061} \\
 & Tail & 0.2149 & 0.2234 & \textbf{0.3335} \\
\bottomrule
\end{tabular}
\end{table}

Recommender systems are often affected by popularity bias, where frequently interacted items receive disproportionate exposure compared to less popular items~\cite{abdollahpouri2019managingpopularitybiasrecommender, Klimashevskaia2024}.
We evaluate SRPFN under popularity-based item groups by stratifying test items into Head, Mid, and Tail groups (20/60/20) according to their training-set interaction frequency.
Table~\ref{tab:pop_bias} reports group-wise HR@5 for SRPFN, SASRec, and FEARec across three datasets.

Across all three datasets and all three popularity groups, SRPFN consistently outperforms both baselines.
The improvements are most significant on the Tail items.
Relative to the strongest baseline, SRPFN improves Tail HR@5 by 55.5\% on Beauty, 81.5\% on Sports, and 49.3\% on Yelp, substantially exceeding its corresponding gains on Head items (4.9\%, 10.1\%, and 2.7\%,
respectively).
These results suggest that SRPFN's gains are not solely driven by disproportionately favoring popular items, which would instead concentrate improvements on the Head group.
Rather, exposure to diverse popularity regimes during pretraining appears to particularly benefit Tail items, for which the target domain signal is sparse.

\subsection{Computational Cost Analysis}

\begin{table}[t]
\centering
\caption{Per-dataset computational cost on a single NVIDIA A6000 GPU, averaged over the Sports and Yelp datasets. For SRPFN, pretraining is performed once offline and takes approximately 15 hours.}
\label{tab:efficiency}
\small
\begin{tabular}{lccc}
\toprule
Computation & SASRec$^\dagger$ & FEARec$^\dagger$ & SRPFN \\
\midrule
Training          & 9 min & 15 h & -- \\
Preprocess  & --     & --    & 14 s \\
Inference         & 35 s   & 179 s & 45 s \\
\cmidrule(lr){1-4}
Total (per dataset) & ${\sim}$10 min & ${\sim}$15 h 
                     & \textbf{${\sim}$1 min} \\
\bottomrule
\end{tabular}
\end{table}

Table~\ref{tab:efficiency} compares the per-dataset computational cost of SRPFN against two representative trained-on-target baselines, measured on Sports and Yelp (the two largest benchmarks by number of interactions; see Table~\ref{tab:dataset_stats} in Appendix~\ref{app:dataset_stats}).
We select SASRec and FEARec as representatives because SASRec is a widely used Transformer-based sequential recommender and shows competitive performance on LastFM in our main results, and FEARec is the overall best-performing trained-on-target baseline across the remaining benchmarks, consistently achieving the closest performance to SRPFN.

Both baselines require gradient-based training for each new dataset, taking approximately 10 minutes and 15 hours, respectively.
In contrast, SRPFN requires no dataset-specific training. It requires preprocessing that computes the low-rank representations (Section~\ref{sec:methods:rep}) and samples support sets, which takes 14 seconds, followed by a single forward pass of 45 seconds.
This implies that after one-time pretraining investment, SRPFN can be applied to new datasets at minimal marginal cost, while achieving competitive or superior performance (Table~\ref{tab:main}).

\subsection{Support Set Analysis}
\label{sec:support_analysis}

\begin{table}[t]
\centering
\caption{Support set size ablation (HR@1).
Columns correspond to $|D_{\mathrm{support}}| \in \{0,1,4,16\}$,
using transition-based selection.}
\label{tab:support_ablation}
\small
\begin{tabular}{lcccc}
\toprule
Dataset & 0 & 1 & 4 & 16 \\
\midrule
Beauty & 0.1185 & 0.2247 & 0.2352 & 0.2364 \\
Yelp   & 0.1651 & 0.2728 & 0.2813 & 0.2891 \\
\bottomrule
\end{tabular}
\end{table}

Table~\ref{tab:support_ablation} analyzes the impact of support set size on SRPFN, using transition-based selection that samples support items from the empirical transition neighborhood.
The setting $|D_{\mathrm{support}}|{=}0$ removes the cross-attention evidence and therefore measures the prediction based only on the encoded query sequence. 
Providing a single support interaction ($|D_{\mathrm{support}}|{=}1$) substantially improves performance, with HR@1 increasing from 0.1185 to 0.2247 on Beauty and from 0.1651 to 0.2728 on Yelp. 
This large gain indicates that conditioning on the support set, even with a single example, enables the model to refine its predictions beyond the query sequence alone. 
Increasing the support set size further improves performance. 
On Beauty, HR@1 improves from 0.2247 to 0.2364 when increasing $|D_{\mathrm{support}}|$ from 1 to 16, and on Yelp from 0.2728 to 0.2891.
Overall, SRPFN refines its predictive distribution at inference time by conditioning on the support set, without requiring updates to target domain parameters.

\subsection{Architecture Ablation}
\label{sec:arch_ablation}
\begin{table}[t!]
\centering
\caption{Architecture ablation on Beauty and Toys. The best performance
in each column is in bold.}
\label{tab:arch_ablation}
\small
\begin{tabular}{lcccc}
\toprule
\multirow{2}{*}{Variant}
 & \multicolumn{2}{c}{Beauty} & \multicolumn{2}{c}{Toys} \\
 \cmidrule(lr){2-3}\cmidrule(lr){4-5}
 & HR@5 & NDCG@5 & HR@5 & NDCG@5 \\
\midrule
SRPFN (full)      & \textbf{0.4418} & \textbf{0.3455} & \textbf{0.4305} & \textbf{0.3389} \\
Shared embeddings & 0.4291 & 0.3337 & 0.4193 & 0.3280 \\
Simple addition   & 0.4208 & 0.3269 & 0.4123 & 0.3207 \\
\bottomrule
\end{tabular}
\end{table}

Table~\ref{tab:arch_ablation} examines the contribution of the main encoder components in SRPFN. 
In \emph{shared embeddings}, the source- and destination-side transition embeddings are collapsed into a single averaged representation, removing the source/destination asymmetry, which tests whether asymmetric transition roles are useful for representing item-to-item dynamics. 
In \emph{simple addition}, the gated residual update is replaced with a direct additive update, removing the consistency-based gate that adaptively controls pairwise transition updates.

Removing the source--destination distinction degrades performance, indicating that treating an item differently depending on whether it appears as a transition source or destination is beneficial, consistent 
with the directional nature of sequential recommendation. 
Removing the gate also reduces performance, suggesting that not all consecutive item pairs should contribute equally to the state update, as some transitions may be noisy or less informative. 
Overall, these results show that SRPFN benefits not only from low-rank input representations themselves but also from encoder designs that preserve directional transition structure and modulate local updates.

\section{Conclusion}
In this work, we presented SRPFN, a Prior-data Fitted Network that performs sequential recommendation without any parameter updates on the target domain. 
By pretraining on datasets sampled from our synthetic prior, SRPFN achieves superior or competitive performance relative to models trained directly on target data across five benchmarks, while requiring only approximately one minute of inference per dataset.
These results suggest that SRPFN learns to approximate the posterior predictive distribution over diverse transition patterns, adapting to new datasets by conditioning on a support set.

We note several limitations that suggest directions for future work.
First, our synthetic prior generates sequences via random walks on static graphs with variable-order lags, and thus does not capture temporal dynamics such as time-decayed preferences, session boundaries, or repeat-consumption patterns. Developing priors that encode richer sequential structures is an important next step. 
Second, while we report full-catalog ranking results (Appendix~\ref{app:full_ranking}), a deeper analysis across more diverse domains remains valuable.

\begin{acks}
This work was partly supported by the Institute for Information \& Communications Technology Planning \& Evaluation (IITP) grants funded by the Korean government (MSIT) 
(No. RS-2026-25526850, High-Efficiency Neural Networks for Artificial General Intelligence (HERMES-Net)), 
the InnoCORE program of the Ministry of Science and ICT (AI Meta-Scientist, N10260110),
Samsung Electronics Co., Ltd. (No. G01240136, KAIST Semiconductor Research Fund (2nd))
and the Korea Advanced Institute of Science and Technology (KAIST) grant funded by the Korea government (MSIT) (No. G04240001, Physics-inspired Deep Learning).
J.~Choi was additionally supported by the KAIST Jang Young Sil Fellow Program.
\end{acks}

\clearpage
\bibliographystyle{ACM-Reference-Format}
\balance
\bibliography{refs}

\clearpage

\appendix

\section{Additional Experimental Details}
\label{app:exp_details}

\subsection{Dataset Statistics}
\label{app:dataset_stats}

Table~\ref{tab:dataset_stats} summarizes the five evaluation datasets after preprocessing.
All datasets are filtered with a 5-core threshold, and interactions are sorted chronologically per user.

\begin{table}[!t]
\centering
\caption{Dataset statistics after preprocessing.}
\label{tab:dataset_stats}
\setlength{\tabcolsep}{4.5pt}
\renewcommand{\arraystretch}{1.06}
\small
\resizebox{\columnwidth}{!}{%
\begin{tabular}{lccccc}
\toprule
Dataset & Beauty & Sports & Toys & Yelp & LastFM \\
\midrule
\# Users          & 22{,}363  & 35{,}598  & 19{,}412  & 30{,}431  & 1{,}090 \\
\# Items          & 12{,}101  & 18{,}357  & 11{,}924  & 20{,}033  & 3{,}646 \\
Avg.\ sequence length& 8.9       & 8.3       & 8.6       & 10.4      & 48.2 \\
\# Interactions   & 198{,}502 & 296{,}337 & 167{,}597 & 316{,}354 & 52{,}551 \\
Sparsity          & 99.93\%   & 99.95\%   & 99.93\%   & 99.95\%   & 98.68\% \\
\bottomrule
\end{tabular}
}
\end{table}

\subsection{Baseline Implementations}
\label{app:baselines}

\begin{table}[t]
\centering
\caption{Source code used for trained-on-target baselines.}
\label{tab:baseline_repos}
\setlength{\tabcolsep}{5.0pt}
\renewcommand{\arraystretch}{1.05}
\small
\begin{tabular}{ll}
\toprule
Baseline & Repository / Implementation \\
\midrule
GRU4Rec  & \url{https://github.com/hidasib/GRU4Rec} \\
Caser & \url{https://github.com/graytowne/caser_pytorch} \\
SASRec   & \url{https://github.com/pmixer/SASRec.pytorch} \\
FMLPRec  & \url{https://github.com/Woeee/FMLP-Rec} \\
CoSeRec   & \url{https://github.com/YChen1993/CoSeRec} \\
FEARec   & \url{https://github.com/sudaada/FEARec} \\
\bottomrule
\end{tabular}
\end{table}

For trained-on-target sequential recommenders, we use the official or widely adopted public implementations summarized in Table~\ref{tab:baseline_repos}. 

\begin{itemize}[leftmargin=*, topsep=2pt, itemsep=1.5pt]
  \item \textbf{GRU4Rec}~\cite{Hidasi2015SessionbasedRW}: a recurrent model for session-based and sequential recommendation.
  \item \textbf{Caser}~\cite{10.1145/3159652.3159656}: a convolutional sequential model that captures high-order patterns via horizontal and vertical filters on the interaction embedding matrix.
  \item \textbf{SASRec}~\cite{Kang2018SelfAttentiveSR}: a Transformer-based sequential recommender with self-attention over user histories.
  \item \textbf{FMLPRec}~\cite{zhou2022fmlprec}: an MLP-based sequential recommender without attention.
  \item \textbf{CoSeRec}~\cite{liu2021coserec}: a contrastive self-supervised sequential recommender with correlation-aware sequence augmentations.
  \item \textbf{FEARec}~\cite{du2023fearec}: a sequential recommender that incorporates frequency-domain signals through hybrid attention.
\end{itemize}

Training-free baselines are computed directly from the target training split without parameter learning or fine-tuning.

\begin{itemize}[leftmargin=*, topsep=2pt, itemsep=2pt]
  \item \textbf{PopRec}: ranks items by their interaction frequency in the training split.

  \item \textbf{Markov-1}: ranks candidates by empirical first-order transition probabilities
  $\hat{P}(j\mid i)=c(i,j)/\sum_{j'}c(i,j')$, where $c(i,j)$ counts consecutive transitions from $i$ to $j$.

  \item \textbf{Co-Occur}: ranks items by their co-occurrence counts with the user's most recent item,
  ignoring adjacency and order within the user history.

  \item \textbf{ItemKNN}: ranks candidates by cosine similarity to the user's most recent item,
  using item vectors from the user--item interaction matrix.
\end{itemize}

\begin{table}[!h]
\centering
\caption{Full-catalog ranking results across five datasets. The best and second-best results in each row are denoted in bold and underlined, respectively.}
\label{tab:full_ranking}
\begin{tabular}{llrrr}
\toprule
Dataset & Metric & SASRec & FEARec & SRPFN \\
\midrule
\multirow{4}{*}{Beauty}
  & HR@10    & 0.0682 & \textbf{0.0813} & \underline{0.0761} \\
  & NDCG@10  & \underline{0.0383} & \textbf{0.0402} & 0.0361 \\
  & HR@20    & 0.1003 & \underline{0.1132} & \textbf{0.1206} \\
  & NDCG@20  & 0.0464 & \textbf{0.0482} & \underline{0.0472} \\
\cmidrule(lr){1-5}
\multirow{4}{*}{Sports}
  & HR@10    & 0.0334 & \underline{0.0488} & \textbf{0.0503} \\
  & NDCG@10  & 0.0180 & \underline{0.0231} & \textbf{0.0245} \\
  & HR@20    & 0.0499 & \underline{0.0704} & \textbf{0.0784} \\
  & NDCG@20  & 0.0221 & \underline{0.0286} & \textbf{0.0316} \\
\cmidrule(lr){1-5}
\multirow{4}{*}{Toys}
  & HR@10    & 0.0744 & \textbf{0.0920} & \underline{0.0872} \\
  & NDCG@10  & \underline{0.0431} & \textbf{0.0456} & 0.0428 \\
  & HR@20    & 0.1011 & \underline{0.1251} & \textbf{0.1305} \\
  & NDCG@20  & 0.0498 & \underline{0.0539} & \textbf{0.0540} \\
\cmidrule(lr){1-5}
\multirow{4}{*}{Yelp}
  & HR@10    & 0.0276 & \textbf{0.0573} & \underline{0.0566} \\
  & NDCG@10  & 0.0140 & \textbf{0.0367} & \underline{0.0283} \\
  & HR@20    & 0.0469 & \underline{0.0845} & \textbf{0.0914} \\
  & NDCG@20  & 0.0189 & \textbf{0.0435} & \underline{0.0371} \\
\cmidrule(lr){1-5}
\multirow{4}{*}{LastFM}
  & HR@10    & \textbf{0.0771} & 0.0532 & \underline{0.0596} \\
  & NDCG@10  & \textbf{0.0429} & \underline{0.0292} & 0.0271 \\
  & HR@20    & \textbf{0.1110} & 0.0817 & \underline{0.0927} \\
  & NDCG@20  & \textbf{0.0513} & \underline{0.0363} & 0.0352 \\
\bottomrule
\end{tabular}
\end{table}

\section{Full-Catalog Ranking Results}
\label{app:full_ranking}

As shown in Table~\ref{tab:full_ranking}, SRPFN ranks first on all metrics on Sports, and achieves the best HR@20 on four of the five datasets (Beauty, Sports, Toys, and Yelp). On the remaining metrics, target-trained baselines remain competitive, particularly on NDCG@10, where dataset-specific training can provide an advantage in fine-grained full-catalog ranking. These results indicate that SRPFN retains competitive full-catalog ranking performance despite using no target domain gradient updates.

\section*{Ethical Considerations}
This work studies sequential recommendation from an algorithmic perspective and does not introduce new data collection or user-facing systems.
All experiments are conducted on publicly available datasets that have been widely used in prior work~\cite{10.1145/3340531.3411954, zhou2022fmlprec}.
Our approach does not rely on sensitive user attributes, nor does it involve personal data beyond anonymized interaction logs.

\section*{Reproducibility Statement}
For implementing the baselines, we followed the official source codes published by authors as detailed in Table~\ref{tab:baseline_repos}.
Also, we make the source code and pretrained model checkpoint publicly available at \url{https://github.com/woooosung/SRPFN}, with an archival version at \url{https://doi.org/10.5281/zenodo.20403804}.

\end{document}